\documentclass[revtex]{emulateapj}

\begin{document}

\shorttitle{Microvariability of Galactic Bulge Miras}
\shortauthors{Wo\'zniak et al.}

\title{Limits on $I$-band microvariability of the Galactic Bulge Miras}

\author{
P. R. Wo\'zniak,
K. E. McGowan,
and W. T. Vestrand
}

\affil{Los Alamos National Laboratory, MS-D436, Los Alamos, NM 87545}
\email{wozniak@lanl.gov, mcgowan@lanl.gov, vestrand@lanl.gov}

\begin{abstract}

We search for microvariability in a sample of 485 Mira variables with high quality $I$-band light
curves from the second generation Optical Gravitational Lensing Experiment (OGLE-II). Rapid
variations with amplitudes in the $\sim$0.2--1.1 mag range lasting hours to days were discovered
in Hipparcos data by de Laverny et al. (1998). Our search is primarily sensitive to events with
time-scales of $\sim$1 day, but retains a few percent efficiency (per object) for detecting
unresolved microvariability events as short as 2 hours. We do not detect any candidate events.
Assuming that the distribution of the event time profiles is identical to that from the Hipparcos
light curves we derive the 95\% confidence level upper limit of 0.038 yr$^{-1}$ star$^{-1}$ for
the rate of such events (1 per 26 years per average object of the ensemble). The high event rates
of the order of $\sim$1 yr$^{-1}$ star$^{-1}$ implied by the Hipparcos study in the $H_P$ band are
excluded with high confidence by the OGLE-II data in the $I$ band. Our non-detection could still be
explained by much redder spectral response of the $I$ filter compared to the $H_P$ band or by
population differences between the bulge and the solar neighborhood. In any case, the OGLE-II
$I$-band data provide the first limit on the rate of the postulated microvariability events
in Mira stars and offer new quantitative constraints on their properties. Similar limits
are obtained for other pulse shapes and a range of the assumed time-scales and size-frequency
distributions.

\end{abstract}


\section{Introduction}
\label{sec:intro}

Mira variables are a subclass of Red Variables, also known as Long Period Variables (LPVs). They
are low mass AGB giants, mostly with M type spectra (e.g. Gautschy \& Saio 1996). Their large
amplitude light variations on time scales of $\sim$1 year are thought to arise due to the
pulsation instability (e.g. Keeley 1970). There are intriguing reports of rapid light variations
in Mira stars on time scales of a few days (Maffei \& Tosti 1995) or even hours and minutes
(Schaefer 1991) with amplitudes in the range 0.1--1.4 mag. The most systematic study of
microvariability in Mira stars is based on Hipparcos data (de Laverny et al. 1998, hereafter
dL98). It reports the discovery of 51 events in 39 out of 239 observed Mira stars. The amplitudes
of events detected by Hipparcos are between 0.2 and 1.1 mag. To best of our knowledge, so far there
are no published estimates for the rate of such events. Other reports of rapid variations in broad
band photometry of specific Mira variables include Smak \& Wing (1979) or more recently de Laverny
et al. (1997) and Teets \& Henson (2003). From existing data very little can be inferred about the
spectral properties of those rapid variations, but microvariability in several spectral features
has been reported (Odell et al. 1970, Kovar et al. 1972). On the other hand, Smith et al. (2002)
do not detect any rapid variations in the COBE DIRBE mid-infrared light curves of 38 Miras.

Willson \& Struck (2002) explored possible interpretations and favored one involving hot flashes
due to interaction of the extended Mira atmosphere with a Jovian planet or a brown dwarf. This
process was studied numerically by Struck, Cohanim \& Willson (2002, 2004). The scenario can
explain the time scale of the observed microvariability and produces energies and luminosities
roughly of the right order of magnitude, but it ignores the fact that dL98 have seen approximately
equal numbers of positive and negative flux excursions.

Here we use a large set of high quality $I$-band photometry from the Optical Gravitational Lensing
Experiment (OGLE) to constrain the rate of microvariability for Mira stars found in the Galactic
Bulge.

\section{Data and sample selection}
\label{sec:data}

\subsection{OGLE-II light curves}
\label{sec:lightcurves}

We use the photometric data collected, processed and published by the second generation OGLE
experiment (OGLE-II; Udalski, Kubiak, \& Szyma\'nski 1997). Our analysis is based on photometry
from Difference Image Analysis (DIA) of the images collected during observing seasons 1997--1999
(Wo\'zniak et al. 2002, Wo\'zniak 2000). The DIA technique relieves numerous complications
associated with crowded fields and enabled routine photometry as accurate as 0.5\% for bright
unsaturated stars like the brightest AGB variables in the Galactic Bulge. This data set contains
$I$-band light curves for approximately 2$\times10^5$ candidate variable objects with the average
time sampling of once every 1--3 days and typically 250 measurements per object. We defer
a detailed discussion of the time sampling to Section~\ref{sec:signature}. Observations span
the magnitude range between 10.5 and 19.5. The total time baseline of the data set is 3 years with
2 inter-season gaps, each lasting about 4 months.

\subsection{Selecting Mira variables}
\label{sec:selection}

The General Catalog of Variable Stars (GCVS; Kholopov 1998) states that Mira variables have
$V$-band light amplitudes between 2.5 and 11 mag, well pronounced periodicity and periods ranging
from 80 to 1000 days. The amplitudes of Miras show a decreasing trend toward the red and the
near-infrared part of the spectrum. The histogram of amplitudes for LPVs with log(P) larger than
2.0 is bimodal (Mattei et al. 1997) with reasonably good separation between lower amplitude
semi-regular variables (SR) and Miras (M).

\begin{figure}
\epsscale{0.75}
\plotone{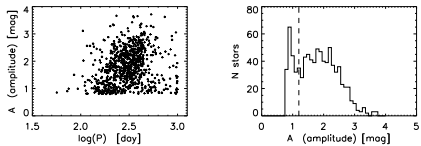}
\caption{Period amplitude distribution (left) and histogram of amplitudes (right) for a sample of
red variables from OGLE-II data. The location of the vertical line indicates the adopted amplitude
threshold for Mira variables.\label{fig:pa}}
\end{figure}

We begin by calculating the total amplitude $A$ for all 2$\times10^5$ objects in the OGLE-II
database of DIA photometry using a running median filter with 11 points and taking the difference
between extreme values. A sample dominated by LPVs is obtained by requiring $A>0.8$ mag. This
selects 842 stars variable on time scales of 2 weeks or longer. Periods are obtained using the
Analysis of Variance algorithm (AOV; Schwarzenberg-Czerny 1989). The period-amplitude diagram and
the histogram of amplitudes for all 842 stars is shown in Figure~\ref{fig:pa}. The gap separating
semi-regulars and Miras is clearly visible at $A\simeq1.2$ mag (c.f. Cioni et al. 2001 and
Eggen 1975). As an independent cross-check, we examine near-infrared colors of candidate stars from
the Two Micron All Sky Survey (2MASS\footnotemark). We require a positive identification with a
2MASS point source object within a 1\arcsec ~radius of the OGLE position. Almost all identified
stars have $(J-H)$ and $(H-K_s)$ colors consistent with AGB variables observed through the reddening
screen equivalent to $A_{_{V}}\simeq$ 0--2 mag of visual extinction. We reject only 3 out of 705
positively identified stars, those having $(J-H)<0.75$ and $(H-K_s)<0.3$. After verifying the 2MASS
colors we select a sample of Mira variables using amplitude ($A>1.2$ mag) and period
(${\rm P}>100$ days). 

\footnotetext{This publication makes use of data products from the Two Micron All Sky Survey, which
is a joint project of the University of Massachusetts and the Infrared Processing and Analysis
Center/California Institute of Technology, funded by the National Aeronautics and Space Administration
and the National Science Foundation.}

\subsection{Quality thresholding}
\label{sec:threshold}

Our analysis is limited to the highest quality data, i.e. light curves with the faintest measurement
of $I<15.0$. In this magnitude range the photon noise is not a dominant part of the error budget
and the influence of crowding is relatively small. Further, only measurements without processing flags
are accepted to avoid possible problems due to saturation, unreliable PSF fits, exceptionally large
seeing FWHM etc. Finally, measurements with unusually large error bars are rejected. The median
and the standard deviation $\sigma$ around the median are calculated for all error bars in a given
light curve. Points with an error bar more than 4$\sigma$ larger than the median error are removed
from the analysis. The latter cut removes only 465 points out of 105,890 and has no influence on
our conclusions (Section~\ref{sec:signature}). The final sample consists of 105,425 $I$-band
measurements of 485 Mira type variables.

\section{Method}
\label{sec:method}

\subsection{Signature of microvariability}
\label{sec:signature}

If microvariability events by tenths of a magnitude or more with durations larger than $\sim$0.1 days
are present, then they should be visible as a tail of photometric outliers. Some of the outliers
could be statistical fluctuations in the measurements, so by assuming that all such outliers are due
to rapid variability in the sources we obtain an upper limit on the frequency of such events.
One can only afford such a conservative assumption and still derive a useful limit when the data quality
is very high and the contribution from random measurement errors to the tail is negligible.

To define the normal behavior of an object with respect to which we measure the outliers, we fit
a smooth curve to the data. Using $\chi^2$ minimization, each season for the light curves is
fitted separately with a model consisting of a set of harmonics:

\begin{equation}
\label{eq:model}
m_I(t) = A_0 + \sum_{k=1}^{7}{\left( A_k \sin(k \omega t) + B_k \cos(k \omega t) \right)},
\end{equation}

\noindent
where $\omega = 2\pi$ yr$^{-1}$. This model is smooth on time scales of 26 days or longer. The
choice of any particular model is not important as long as it is smooth on time scales of about 1
month and still fits the long term behavior of the Mira flux. Oscillating functions can usually
achieve that with fewer parameters than other basis functions.

\begin{figure*}
\epsscale{0.75}
\plotone{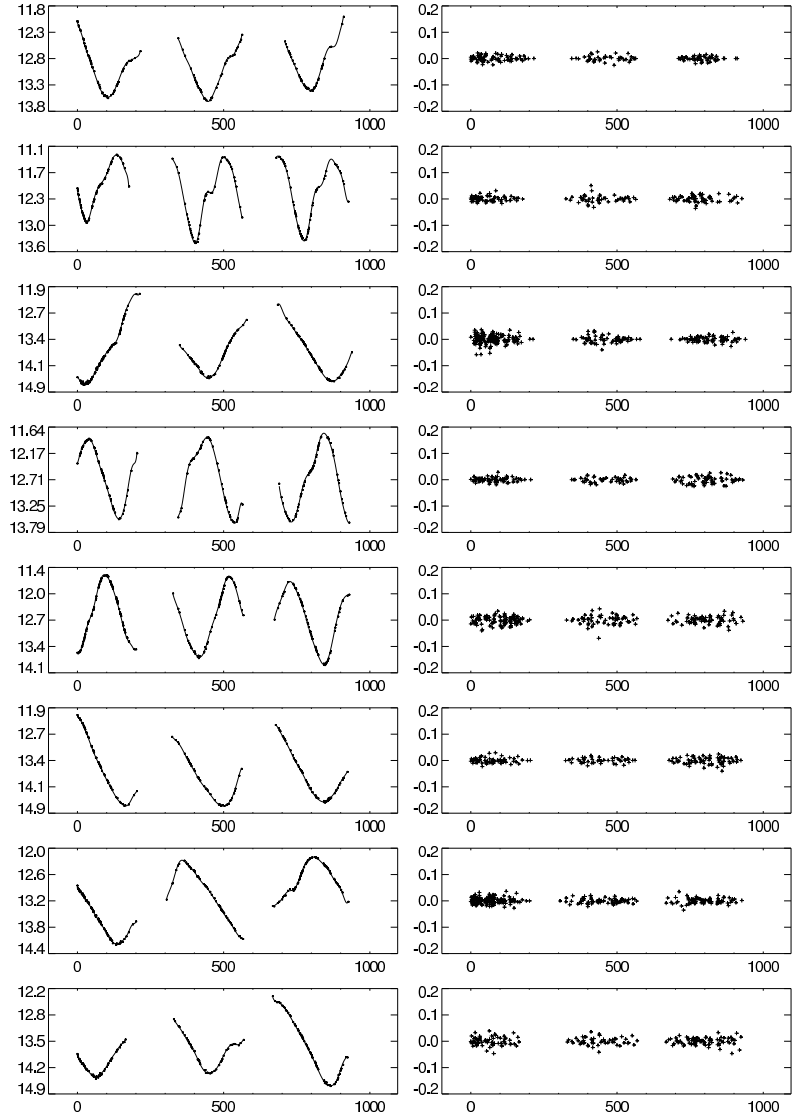}
\caption{Examples of OGLE-II light curves for Mira variables in the Galactic Bulge with model fits
given by Equation~\ref{eq:model} (left). Also shown are photometric residuals with respect to the
model (right). All units are $I$ magnitude versus time in days. Note that the amplitudes of all
events reported by dL98 exceed the largest residuals by a factor of $\sim$5.
\label{fig:examples}}
\end{figure*}

It is important to verify that the
model does not over-fit the data, as this would suppress the detection of outliers.
In Figure~\ref{fig:examples} we present a few examples of OGLE-II light curves for Mira stars used
in our calculations. There is some limited potential for over-fitting near the beginning and the
end of each observing season, especially if one or two points are somewhat separated from the rest
of the measurements in a particular season. This is unavoidable as the model for normal behavior
of the flux near such loose points is effectively undefined. But within the observing season there is
very little possibility that a model smooth on time-scales of 1 month (Figure~\ref{fig:examples})
can fit individual points. The temporal spacing of measurements is too fine for that in most cases.
Nevertheless, in order to make any conclusions about the rate of microvariability, the influence
of this and other possible effects has to be fully quantified with the proper efficiency
calculation. We present such a calculation in Section~\ref{sec:simulations}.

The approach taken by dL98 in the Hipparcos analysis is not possible for OGLE-II data. Hipparcos
light curves consist of brief observing sequences lasting typically less than
1 day but densely sampled with measurements often separated by about 20 minutes. This is why long
term variations of the main Mira cycle are not an issue and at the same time many events can be
resolved. The time sampling of the OGLE-II light curves is generally not sufficient to resolve any
rapid variability on time-scales shorter than 1 day. Typically the density of measurements well
within the observing season corresponds to one point every 1 to 3 days and it tapers off near both
ends of the season. Some variations of temporal sampling arise due to changing observing
conditions. Still, there are 1046 pairs and 210 triples of observations spanning less than 24
hours. Those better sampled regions significantly improve the sensitivity to the shortest events
considered in our analysis. At the basis of our approach is exceptionally good error distribution
in the OGLE-II DIA photometry combined with the ability to fit longer term behavior of the Mira flux.
Even if microvariability events are infrequent and very short, each object with rapid variations
discovered by dL98 spends some (small) fraction of time in the events, when the object flux
is far from the normal smooth trend of the main pulsation cycle. Eventually, with persistent
monitoring, some measurements will be affected by microvariability regardless of the time sampling.

We define an ``event'' as one or more consecutive measurements deviating (consistently) up or
down by more than 0.2 mag with respect to the best fit model of Equation~\ref{eq:model}. Experiments
requiring two consecutive deviations (a two-point filter) and 0.15 mag limit for the flux departure
generally confirm our conclusions, albeit at a factor of 2 to 10 lower sensitivity. In both searches
we detect no events.

We checked the impact of the last selection cut from Section~\ref{sec:threshold} which removes
points with unusually large error bars. Without this condition imposed, only three single point
``events'' in the entire data set depart from the model by: 0.21, $-$0.25, 0.23 mag. Two of them
originate from the same frame, and all have error bars much larger than the rest of the
corresponding light curve, particularly the neighboring points. They are much more likely
manifestations of a problem with the measurement rather than detections of microvariability. It is
safe to conclude that we have not detected any rapid variability events.

\subsection{Probability density for event rate}
\label{sec:probability}

Counting events from Section~\ref{sec:signature} is a Poisson process with the expectation value
$\mu=\sum_s \nu T \epsilon_s = \nu \mathcal{E}$, where $\nu$ is the yearly rate of actual physical
events per object of the ensemble, $T$ is the duration of the survey and $\epsilon_s$ is the
efficiency of detecting an event that occurred randomly during the survey time in star $s$. We
substitute $\mathcal{E}=T\sum_s \epsilon_s$ to simplify notation. In other words $\mu$ is the
total expected number of events in the entire sample of 485 stars observed for $T=3$ years
assuming that every star undergoes $\nu$ events per year on average. Equivalently, if only a
fraction of the stars experienced events, the rate $\nu$ would be proportionately higher in those
objects. Using Bayes' theorem we can derive the probability distribution for the rate given the
number of observed events $n$: 

\begin{equation}
\label{eq:Bayes}
P(\nu|n) = {{P(n|\nu) P(\nu)}  \over { P(n)}},
\end{equation}

\noindent
where $P(n|\nu)=\mu^n \exp(-\mu)/n!$ is the Poisson distribution and $P(n)$ is the normalization
factor obtained from condition $\int_{0}^{\infty}P(\nu|n)d\nu = 1$. For lack of other knowledge we
also assume a uniform prior probability for the rate $P(\nu) \equiv 1$. This results in the
following probability distribution for $\nu$:

\begin{equation}
\label{eq:prob}
P(\nu|n) = {{\mathcal{E}^{n+1}}\over{n!}} \nu^n \exp(-\nu \mathcal{E}).
\end{equation}

\noindent
In the case of a non-detection $P(\nu|n=0) = \mathcal{E}\exp(-\nu \mathcal{E})$ which can be
integrated for a confidence level $\alpha$ to obtain the upper limit for the rate:

\begin{equation}
\label{eq:R95}
R_{\alpha} = {{-\ln(1-\alpha)} \over {\mathcal{E}}}.
\end{equation}

\noindent
In the following discussions we will use $R_{95}$, the 95\% confidence upper limit for the event
rate.

\subsection{Simulating efficiency}
\label{sec:simulations}

The term $\mathcal{E}$ is the crucial piece of information and can be evaluated with the use of a
Monte Carlo simulation. We consider microvariability events with two kinds of flux profiles:
triangular and instantaneous rise with exponential decay (Sections~\ref{sec:deLaverny} and
\ref{sec:flashes} respectively). The time-scale ($\tau$) and the amplitude of a simulated event
($\Delta m$) are drawn from their assumed distributions. A random epoch within the three-year
survey baseline is also generated and the flux variation is injected at that epoch. Finally, we
attempt to detect the event using the same procedure employed with the real observations
in Section~\ref{sec:signature}. This includes re-fitting the empirical curve describing normal
smooth behavior. For each of the 485 stars in the ensemble, we generate 1000 trial events
and count all successful detections.

\section{Results}
\label{sec:results}

\subsection{Derived rate of Hipparcos events}
\label{sec:deLaverny}

\begin{figure}
\epsscale{0.75}
\plotone{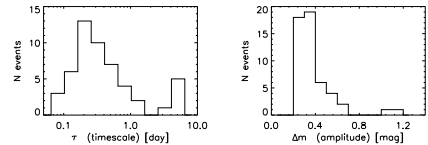}
\caption{Distribution of time scales (left) and amplitudes (right) of events reported by
de Laverny et al. (1998).\label{fig:deLaverny}}
\end{figure}

The distributions of time-scales and amplitudes for microvariability events reported by dL98 are
shown in Figure~\ref{fig:deLaverny}. For the purpose of constraining the rates of these types of
event, the simulated event time-scales and amplitudes are drawn from the actual observed
distributions. There were approximately as many upward and downward deviations reported in the
Hipparcos data. We verified that the two distributions appear uncorrelated. In all our simulations
we assume that the event amplitude is independent of the time-scale and the sign of the deviation
is random. The shape of the artificial impulse is a triangle with the base covering the time
interval $\pm \tau$. In Section~\ref{sec:flashes} we will see that the results are not very
sensitive to that choice.

\begin{figure}
\epsscale{0.75}
\plotone{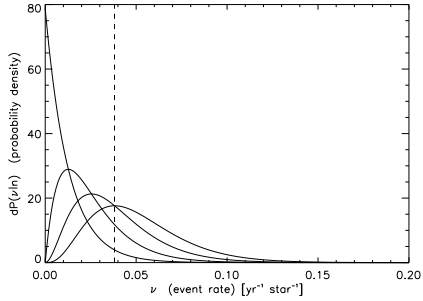}
\caption{Probability distributions for the rate of events of the de Laverny et al. (1998) type.
Lines correspond to probabilities given that there were between $n$ = 0 (farthest left) to $n$ = 3
(farthest right) detections. The location of the vertical line indicates the 95\% confidence upper
limit for the rate given a non-detection.\label{fig:rate}}
\end{figure}

Figure~\ref{fig:rate} shows the probability distributions for the event rate $\nu$ given
$n$ = 0, 1, 2, and 3 detections. Given that we have detected $n$ = 0 events in the OGLE-II data,
the upper limit for the rate (95\% confidence) is $R_{95}$ = 0.038 yr$^{-1}$ per star. We conclude,
with 95\% probability, that dL98 type events are less frequent than 1 every 26 years in any given
object of our OGLE $I$-band sample.

The study of dL98 does not give the value for the efficiency of the Hipparcos light curves in
detecting Mira microvariability. In order to derive the event rate implied by results of dL98 we
need to know the detection efficiency. We can estimate it using public Hipparcos data from the Epoch
Photometry Annex (ESA 1997). For any given object the Hipparcos observing sequences cover only a
small fraction of the total duration of the experiment. Observing sequences are typically separated
by gaps lasting about two weeks. Any events occurring in those gaps cannot be detected; the instrument
is simply not observing the object. We extracted light curves for 226 Hipparcos stars classified as
Miras in GCVS. This is fewer than 239 light curves analyzed by dL98, but includes objects
contributing 50 out of 51 reported events\footnotemark. We adopt 3.37 years as the total duration
of the Hipparcos mission, taking the difference between the first and the last epoch in all 226 light
curves. For each light curve we subtract from the total duration the sum of all gaps lasting 10 days
or longer and divide the result by the total duration. On average there were 104 photometric points
available per star. We do not remove any flagged points. This and the implicit assumption of 100\%
efficiency within observing sequences will only increase the final value of the efficiency and lower
the rate.

\footnotetext{There seems to be a problem with identification of the 51st event in
variable CE Lyr. The on-line Hipparcos catalog entry for the star closest to the GCVS position
states that the star may be wrongly identified with CE Lyr. The star appears constant. In any case,
these minor differences are unimportant.}

\begin{figure}
\epsscale{0.75}
\plotone{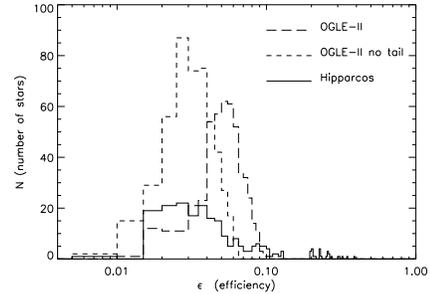}
\caption{Results of the Monte Carlo efficiency experiments with OGLE-II data compared to a
conservative estimate of the efficiency for Hipparcos data. The thin short-dash line is for the
Monte Carlo run on OGLE-II data using the distribution of time-scales from dL98 truncated above
$\tau = 2$ days.\label{fig:eff_distr}}
\end{figure}

The distribution of the resulting efficiencies for Hipparcos data is shown in
Figure~\ref{fig:eff_distr} and compared to the results of our Monte Carlo experiments with OGLE-II
data. For a few Hipparcos stars events can be detected with relatively high efficiency reaching
20--40\%. However, for the majority of the Hipparcos light curves the efficiency is similar to that
for OGLE-II data. The additional Monte Carlo run on OGLE-II data with the distribution of
time-scales from dL98 truncated at $\tau > 2$ days demonstrates that our results do not
depend dramatically on the presence of the long duration tail.

If $\nu$ is the true yearly rate of events in units of yr$^{-1}$ star$^{-1}$, the expected number of
events over the total duration of the experiment (detected or not) is: $\nu N T$, where $N$ is the
total number of monitored objects and $T$ is the duration of the experiment in years. On the other
hand, the total number of events can be recovered from the number of detected events by counting
each detection with the weight proportional to the inverse of the efficiency:
$\sum_s n_s/\epsilon_s$, where $n_s$ is the number of detected events in star $s$ and $\epsilon_s$
is the efficiency of detecting a random event in star $s$. The number of detections per star in
dL98 study is between 0 and 3. By equating the two above expressions one obtains an estimator
of the event rate:

\begin{equation}
\label{eq:rate}
\nu = {{1} \over {N T}} \sum_s {{n_s} \over {\epsilon_s}}
\end{equation}

We evaluate the rate in Equation~\ref{eq:rate} using our efficiency estimates for a sample of $N = 226$
Hipparcos light curves and accepting all detections reported by dL98 (except the one in CE Lyr for
which we could not find the data). As mentioned earlier, the efficiency was likely over-estimated
so the rate is likely under-estimated. The resulting rate is 1.63 yr$^{-1}$ star$^{-1}$. Therefore,
the Hipparcos results imply that microvariability events in Miras are quite common in the spectral
region covered by the $H_P$ band. This is due to the fact that, assuming the events are not correlated
with the scheduling of the Hipparcos observing sequences, for every detected event there were $\sim$25
events that could not be detected due to a gap in coverage. The effective efficiency is only $\sim$4\%
and comparable to that of the OGLE-II observations. Taken at face value, the OGLE-II data appear
inconsistent with high rates implied by the number of Hipparcos detections. Given a true rate of 1 event
per year per object, the probability that an ensemble like ours would yield 0 detections is less than
$6\times10^{-35}$. However, there are several scenarios that can easily
accommodate both our non-detection in the $I$ filter and the positive Hipparcos result in the $H_P$
band. We discuss this issue further in Section~\ref{sec:discussion}.

\vspace{1cm}

\subsection{Constraining flashes on Miras}
\label{sec:flashes}

\begin{figure}
\epsscale{0.75}
\plotone{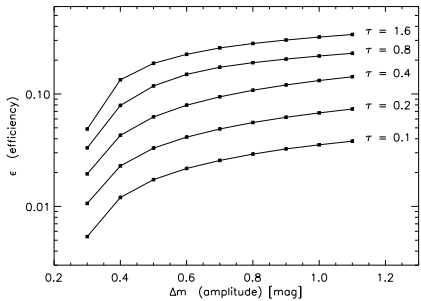}
\caption{Simulated efficiency of detecting flashes with instantaneous rise and exponential decay.
The dependence on amplitude $\Delta m$ is given for several fixed time scales $\tau$.
\label{fig:efficiency}}
\end{figure}

Several possible mechanisms for rapid light variations in Mira variables have been discussed by
Willson and Struck (2002). One such scenario invokes a sudden release of energy leading to "hot
flashes". We can put useful constraints on such events by simply simulating the efficiency of
detecting a different kind of flux impulse in our data. This also provides a cross-check on the
results from Section~\ref{sec:deLaverny}, and in particular shows how sensitive the derived limits
on frequency are with respect to the pulse shape. In this section we adopt a flare with instantaneous
rise by $\Delta m$ and exponential decay time $\tau$. The average efficiency (per
star) of detecting such flares is shown in Figure~\ref{fig:efficiency}. The dependence on the flare
amplitude between 0.3 and 1.1 mag is plotted for a number of time scales $\tau$ ranging from 0.1
to 1.6 days. Note that the figure shows that flashes with amplitudes of a few tenths of a magnitude
lasting only a couple hours can be detected with the efficiency of a few percent in any of the stars
in our ensemble.

Before deriving the upper limits on rates, we introduce the size-frequency distribution into this
calculation. The published data offers very little information on the properties of suspected
flares, but one generally expects strong events to occur less frequently and weak events to
dominate. For the purpose of illustrating the range of possible outcomes, we assume a power-law
distribution of $f = \Delta F/F$ the amplitude of the simulated flashes relative to the flux $F$
of the parent star:

\begin{equation}
\label{eq:powerlaw}
N(f)d f\propto f^{-\gamma} d f
\end{equation}

\noindent
At fixed time scale $\tau$ and parent star flux $F$ this size distribution is equivalent to assuming
a power-law distribution of flare energies. The distribution is truncated below the relative flare
amplitude of 20\% and above 100\%, which corresponds to amplitudes of about 0.2 and 0.75 mag
respectively.

\begin{figure}
\epsscale{0.75}
\plotone{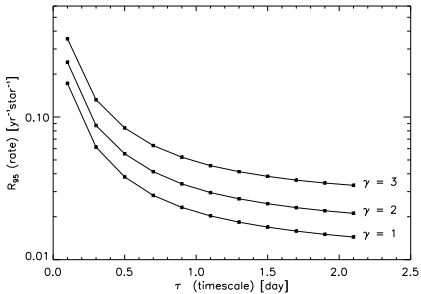}
\caption{Upper limits (95\% confidence) on the rate of flashes on Mira variables given a
non-detection in our ensemble. The dependence on time scale is given for three power law
distributions of fractional flux amplitudes with the index $\gamma$ = 1, 2, and 3
(Section~\ref{sec:flashes}).\label{fig:limits}}
\end{figure}

\noindent
Taking the amplitude distribution of dL98 and ignoring the sign of the variation, $\gamma$ between
2.0 and 2.5 gives a good approximation within the limited statistics of the data set. We evaluate
the efficiency at several fixed time scales ranging from 0.1 to 2.0 days for a power-law size
frequency distribution with index $\gamma = $1, 2 and 3. The resulting rates are given in
Figure~\ref{fig:limits}. The frequency of 1 per 2.5 years in any given star is excluded with 95\%
probability for all types of considered events down to durations of 2.4 hours. The limits quickly
tighten for longer flare durations. For $\gamma$ = 2 we can exclude 0.5 day long events being more
frequent than about once every 17 years.

\section{Discussion}
\label{sec:discussion}

Using a large sample of objects with high quality light curves we place stringent limits on the
frequency of rapid variability events in the Galactic Bulge Mira variables in the $I$ band. To the
best of our knowledge this is the first published limit for the rate of rapid variability in
Miras. The study is most sensitive to events lasting about 1 day or longer, however it retains
a few percent efficiency (per star) for detecting unresolved microvariability on time scales as
short as 2 hours. With 95\% probability we excluded events of the type found by dL98 occurring
more frequently than 1 per 26 years in the average object of our sample
(0.038 yr$^{-1}$ star$^{-1}$). The results are not sensitive to the details of the temporal
profile of the flares. The most significant dependence of the detection efficiency in our ensemble
is on the event time scale because of the 1--3 day light curve sampling. The limits on the
frequency of flashes with a sudden flux rise and exponential fall-off with power-law distribution
of energies (at fixed parent star luminosity) have been presented. For flare durations ranging
from 0.1 to 1 day the corresponding limiting rates are between once every few years and once every
few tens of years.

In our calculations we essentially extrapolate the properties of the rapid variability events
observed by Hipparcos from $H_P$ band to the $I$ band. Under that assumption, the OGLE-II data
are inconsistent with high rates implied by the number of Hipparcos detections. A possible explanation
is that the microvariability is much less prominent or the events are shorter in the $I$ band. In that
case we are still learning something about microvariability in Mira stars. The typical accuracy
of the OGLE-II photometry is much better than 0.2 mag (Figure~\ref{fig:examples}) and we could
not find any indication of rapid variations in any of the 485 studied objects. All light curves
have very smooth appearance.

There are at least three possible explanations for the difference between the two results:

\begin{enumerate}

\item Mira variables in the Galactic Bulge population differ from their counterparts in the solar
neighborhood. In the scenario of a planet interacting with the Mira atmosphere, e.g., the Bulge
population might not have any planets. It is plausible that the Galactic Bulge Miras have lower
metallicity than a typical Hipparcos star. Planets are more prevalent around metal-rich stars
(e.g. Santos, Israelian, \& Mayor 2004). It has been suggested that dust may play an important
role in microvariability of Miras (Stencel et al. 2003). The dust content of the stellar envelope
could be related to metallicity.

\item Microvariability in Miras is occurring in the part of the spectrum that is not covered by
Cousins $I$ band but is contributing to the blue part of the $H_P$ band.

\item Microvariability in Miras is less common than implied by results of dL98. The only sure way
to verify this is to collect more data and carefully analyze it. Numerous groups are setting up
fast cadence photometric monitoring experiments with the goal of detecting real time transients
(Vestrand et al. 2002) or finding transiting Jovian planets (Horne 2003). Teets \& Henson (2003) are
conducting a special purpose monitoring campaign of $\sim$5 sources previously reported to display
rapid variability using $B, V, R, I$ filters. The current data set covers many nights spanning the
time interval of over 2 years with up to 5 photometric points per hour for up to 8 hours a night.
So far only a single event in RR Boo has been detected. With the available phase coverage this is
far below expectations based on the Hipparcos data. 

\end{enumerate}

The $H_P$ filter of the Hipparcos mission is broader than $V$ and still sensitive in the
wavelength range 650--900 nm, where the spectral energy distribution of a Mira variable rises
steeply. We can calculate the amount of energy emitted by a Mira star in the $H_P$ band knowing
that the ratio of luminosities in $H_P$ and $V$ is about 2.5. This is easily estimated assuming
that the spectrum of a Mira star blue-ward of 1000 nm is a black body with effective temperature
$T_{eff} = 2500$ K and by folding it with filter transmission curves (Johnson 1965, van Leeuwen et
al. 1997). Then using the absolute magnitudes $M_V = -0.3$ of an M5 giant and the $V$-band Solar
luminosity $L_{V, \odot} = 4.64\times10^{32}$ erg s$^{-1}$ (Binney \& Merrifield 1998) we obtain
$L_{H_P} = 1.3\times10^{35}$ erg s$^{-1}$ for a Mira star. We adopt $M_{V, \odot}$ = 4.83. Mira
variables vary considerably less in the $H_P$ band compared to $V$ with amplitudes rarely
exceeding 4 mag (Whitelock, Marang \& Feast 2000). Even during minimum light and taking an extreme
value for the amplitude (4 mag) the total energy of a flare with amplitude $\Delta m$ and duration
$\tau$ is:

\begin{equation}
\label{eq:energy}
E = 2.8\times10^{38} {\rm erg} \times C \times \left({{\tau} \over {1 ~\rm day}}\right)\times
\left(10^{0.4 |\Delta m|} - 1\right),
\end{equation}

\noindent
where $C$ is the numerical factor depending on the profile of the light variation. For a top hat
function $C = 1$. We can see that a flare of a third of a magnitude lasting 24 hours will require
$10^{38}$ ergs of energy---an order of magnitude more than the energy required for the same
contrast in the $V$ band calculated by Willson \& Struck (2002). If in fact the events detected by
Hipparcos are confirmed, one will be faced with finding a large energy source that is triggered
approximately as often as an energy sink of the same size. The observational evidence is
accumulating and it should not be long before we can formulate a more definitive answer to the
exciting question of rapid high amplitude variability on Miras.

\acknowledgments

P. Wo\'zniak was supported by the Oppenheimer Fellowship at LANL. Additional support was provided
by the DOE contract W-7405-ENG-36 to the RAPTOR project. This paper is based on observations
obtained with the 1.3-m Warsaw Telescope at the Las Campanas Observatory of the Carnegie
Institution of Washington.


\begin{thebibliography}

\bibitem[Binney(1998)]            {bin98}     Binney, J., \& Merrifield, M. 1998, Galactic Astronomy
                                              (Princeton, NJ: Princeton University Press)

\bibitem[Cioni(2001)]             {cio01}     Cioni, M. R. L., Marquette, J. B., Loup, C., Azzopardi, M.,
                                              Habbing, H. J.,
                                              Lasserre, T., \& Lesquoy, E. 2001, \aap, 377, 945    

\bibitem[deLaverny(1997)]         {deL97}     de Laverny, P., Geoffray, H., Jorda, L., \& Kopp, M.
                                              1997, \aaps, 122, 415

\bibitem[deLaverny(1998)]         {deL98}     de Laverny, P., Mennessier, M. O., Mignard, F., \& Mattei,
                                              J. A. 1998, \aap, 330, 169 (dL98) 

\bibitem[Eggen(1975)]             {egg75}     Eggen, O. J. 1975, \aj, 195, 661

\bibitem[ESA(1997)]               {ESA97}     ESA 1997, The Hipparcos and Tycho Catalogues,
                                              ESA SP-1200

\bibitem[Horne(2003)]             {horne03}   Horne, K. 2003, preprint (astro-ph/0301249)

\bibitem[Johnson(1965)]           {joh65}     Johnson, H. L. 1965, \apj, 141, 923

\bibitem[Keeley(1970)]            {kee70}     Keeley, D. A. 1970, \apj, 161, 657

\bibitem[Kholopov(1998)]          {khol98}    Kholopov, P. N. 1998, General Catalog of Variable Stars
                                              (4th ed.; Moscow: Nauka)

\bibitem[Kovar(1972)]             {kov72}     Kovar, R. P., Potter, A. E., Kovar, N. S., \&
                                              Trafton, L. 1972, \pasp, 84, 46

\bibitem[Mattei(1997)]            {mat97}     Mattei, J. A., Foster, G., Hurwitz, L. A., Malatesta, K. H.,
                                              Willson, L. A., \&
                                              Mennessier, M. O. 1997, in Proc. of the ESA Symposium 402,
                                              Hipparcos-Venice '97,
                                              ed. B. Battrick, M. A. C. Perryman, \& P. L. Bernacca
                                              (ESA Publication Division: Noordwijk, The Netherlands), 269                 

\bibitem[Odell(1970)]             {ode70}     Odell, A., Vrba, F., Fix, J., \& Neff, J. 1970,
                                              \pasp, 82, 883

\bibitem[Santos(2004)]            {san04}     Santos, N. C., Israelian, G., \& Mayor, M. 2004,
                                              \aap, 415, 1153

\bibitem[Schaefer(2000)]          {sch02}     Schaefer, B. E., King, J. R., \& Deliyannis, C. P. 2000,
                                              \apj, 529, 1026

\bibitem[Schwarz(1989)]           {sch89}     Schwarzenberg-Czerny, A. 1989, \mnras, 241, 153

\bibitem[Smak(1979)]              {smak79}    Smak, J., \& Wing, R. F. 1979, Acta Astron., 29, 199 

\bibitem[Smith(2002)]             {smi02}     Smith, B. J., Leisawitz, D., Castelaz, M. W.,
                                              \& Luttermoser, D. 2002, AJ, 123, 948

\bibitem[Stencel(2003)]           {ste03}     Stencel, R. E., Ostrowski-Fukuda, T. A., Jurgenson, C. A.,
                                              \& Phillips, A. 2003, in ASP Conf. Series, Proceedings
                                              of the 12th Cambridge Workshop on Cool Stars,
                                              Stellar Systems and the Sun, eds. A. Brown, G. M. Harper,
                                              \& T. R. Ayres, (San Francisco: ASP), in press

\bibitem[Struck(2002)]            {str02}     Struck, C., Cohanim, B. E., \& Willson, L. A. 2002,
                                              \apjl, 572, L83

\bibitem[Struck(2004)]            {str04}     Struck, C., Cohanim, B. E., \& Willson, L. A. 2004,
                                              \mnras, 347, 173

\bibitem[Teets(2003)]             {tee03}     Teets, W. K., \& Henson, G. D. 2003, \baas, 35, 1217

\bibitem[Udalski(1997)]           {ud97}      Udalski, A., Kubiak, M., \& Szyma\'nski, M. 1997,
                                              Acta Atron., 47, 319

\bibitem[Vestrand(2002)]          {vest02}    Vestrand, W. T., et al. 2002, in Proc. of SPIE, 4845,
                                              Advanced Global Communications Technologies
                                              for Astronomy II,
                                              ed. by R. I. Kibrick, (Bellingham: SPIE), 126

\bibitem[Whitelock(2000)]         {whi00}     Whitelock, P., Marang, F., \& Feast, M. 2000,
                                              \mnras, 319, 728

\bibitem[Willson(2002)]           {wil02}     Willson, L. A., \& Struck, C. 2002, J. AAVSO, 30, 23

\bibitem[Wozniak(2000)]           {woz00}     Wo\'zniak, P. R. 2000, Acta Astron., 50, 421

\bibitem[Wozniak(2002)]           {woz02}     Wo\'zniak, P. R., Udalski, A., Szyma\'nski, M., Kubiak, M.,
                                              Pietrzy\'nski, G., Soszy\'nski, I., \&
                                              \.Zebrun, K. 2002, Acta Astron., 52, 129

\bibitem[vanLeeuwen(1997)]        {vanl97}    van Leeuwen, F., Evans, D. W., Grenon, M., Grossmann, V.,
                                              Mignard, F., \& Perryman, M. A. C. 1997, \aap, 323, 61

\end{thebibliography}
\end{document}